\pdfoutput=1

\documentclass[aps, twocolumn,prd, superscriptaddress, preprintnumbers, floatfix, longbibliography,nofootinbib,10pt]{revtex4-2}
\usepackage{amsfonts,amsmath,amssymb,amstext}
\usepackage{dsfont,bm}
\usepackage[utf8]{inputenc}
\usepackage{newtxtext}
\usepackage[upint]{newtxmath}
\usepackage{microtype}
\usepackage{textcomp}
\usepackage{eucal}
\usepackage{bm}
\usepackage{siunitx}
\usepackage{comment}
\usepackage{lipsum}
\usepackage{mathtools}
\usepackage{tikz}
\usetikzlibrary{calc,decorations.pathmorphing}
\usepackage{enumerate}
\usepackage{amsfonts}
\usepackage{amsmath}
\usepackage{amssymb}
\usepackage{color}
\usepackage{soul}

\usepackage{graphicx}

\usepackage[colorlinks,allcolors=blue]{hyperref}
\usepackage[capitalize]{cleveref}

\definecolor{DarkRed}{rgb}{0.65,0,0}%
\definecolor{Green}{rgb}{0,0.3,0.3}
\definecolor{Purple}{rgb}{0.3,0,0.65}
\definecolor{Red}{rgb}{1,0,0}
\definecolor{Blue}{rgb}{0,0,0.85}
\definecolor{Magenta}{rgb}{1,0,1}

\newcommand{\bo}[1]{\boldsymbol{#1}}

\newcommand{\ve}[1]{\boldsymbol{#1}}
\newcommand{\ca}[2][]{c_{#2}^{\vphantom{\dagger}#1}} % op. c (annihilate) 
\newcommand{\cc}[2][]{c_{#2}^{{\dagger}#1}}          % op. c dagger (create) 
\newcommand{\aaaa}[2][]{\alpha_{#2}^{\vphantom{\dagger}#1}} % op. c (annihilate) 
\newcommand{\ac}[2][]{\alpha_{#2}^{{\dagger}#1}}          % op. c dagger (create) 
\newcommand{\ba}[2][]{\beta_{#2}^{\vphantom{\dagger}#1}} % op. d (annihilate) 
\newcommand{\bc}[2][]{\beta_{#2}^{{\dagger}#1}}       
\newcommand{\aaa}[2][]{a_{#2}^{\vphantom{\dagger}#1}} % op. c (annihilate) 
\newcommand{\aac}[2][]{a_{#2}^{{\dagger}#1}}          % op. c dagger (create) 
\newcommand{\bba}[2][]{b_{#2}^{\vphantom{\dagger}#1}} % op. d (annihilate) 
\newcommand{\bbc}[2][]{b_{#2}^{{\dagger}#1}}  
\newcommand{\vecr}{\ve{r}}

\newcommand{\eg}{\textit{e.g. }}%[syn: f.eks., for example, for instance]
\newcommand{\etal}{\emph{et al.}}

\newcommand{\be}{\begin{equation}}
\newcommand{\ee}{\end{equation}}

\newcommand{\prlsection}[1]{\textit{#1}.\kern0.05em---\kern0.05em\ignorespaces}

\begin{document}
\title{{Spin pumping in an altermagnet/normal metal bilayer}}
\author{Erik Wegner Hodt}
\email[Corresponding author: ]{erik.w.hodt@ntnu.no}
\affiliation{Center for Quantum Spintronics, Department of Physics, Norwegian \\ University of Science and Technology, NO-7491 Trondheim, Norway}
\author{Jacob Linder}
\affiliation{Center for Quantum Spintronics, Department of Physics, Norwegian \\ University of Science and Technology, NO-7491 Trondheim, Norway}

\begin{abstract}
Altermagnetism is a subclass of antiferromagnetism that features spin-polarized electron bands of a non-relativistic origin despite the absence of a net magnetiation in the material.
We here theoretically study spin pumping from an altermagnetic insulator into a normal metal. The symmetry properties of the lattice and spin order of the altermagnet alters the magnon dispersion compared to a conventional square lattice antiferromagnet. We find that for a homogeneous magnetic field, the spin pumping current is the same as that of a regular antiferromagnet. If, however the magnetic field becomes spatially dependent, we predict that the altermagnetic order leaves a unique fingerprint on the spin pumping behaviour when the orientation of the spatial modulation does not align with the high-symmetry paths of magnon degeneracy in the altermagnet. This demonstrates that altermagnets can be used for THz spin pumping purposes with novel behaviour, distinguishing them from their regular antiferromagnet counterparts.
\end{abstract}
\maketitle

\section{Introduction}
{Injecting spins into materials is an important concept in the research field of spintronic devices \cite{hirohata_book_20}. Such injection can be achieved in different ways, such as applying a spin-polarized electric current or via spin pumping. The spin pumping technique \cite{tserkovnyak2002sp, tserkovnyak_rmp_04} consists of setting the spins in a magnetic material into precessional motion, which causes the material to emit a flow of spin into an adjacent material. By varying which material the spins are pumped into, the spin current can be modified depending on the material properties. Moreover, the absorption of the spin current can provide useful information about the band structure and spin-dependent interactions in the material receiving the spin current. However, one can also vary the material from which the spin current is pumped. Spin pumping is possible using both ferromagnetic \cite{Tserkovnyak2002Dec, Tserkovnyak2002Feb} antiferromagnetic materials \cite{cheng_prl_14, takei_prb_15}, metals as well as insulators.}

{Recently, a class of antiferromagnetic materials known as altermagnets has sparked much interest in the research community \cite{ahn_prb_2019, hayami_jpsj_19, smejkal_sa_20, yuan_prb_20}. These materials have features in common with both ferromagnets and antiferromagnets \cite{smejkal_prx_22a, smejkal_prx_22b, mazin2023_altermagnetism}. Similarly to conventional collinear antiferromagnets, they break time-reversal symmetry but have no net magnetization as the time-reversal operation can be nullified by a lattice translation operation. However, in contrast to such materials and in similarity to ferromagnets, altermagnets feature spin-polarized electron bands. This requires $PT$-symmetry breaking where $P$ is the parity operation and $T$ is the time-reversal operation. This also modifies the magnon dispersion relation compared to conventional antiferromagnets \cite{cui_arxiv_23, brekke_arxiv_23}. Predictions for altermagnetic materials span a range of different materials: insulators like FeF$_2$ and MnF$_2$, semiconductors like MnTe, metals like RuO$_2$, to superconductors
like La$_2$CuO$_4$ \cite{smejkal_prx_22a, fedchenko_arxiv_23, smejkal_arxiv_22,hariki_arxiv_23}}.

\begin{figure}[hbt]
    \centering
    \includegraphics{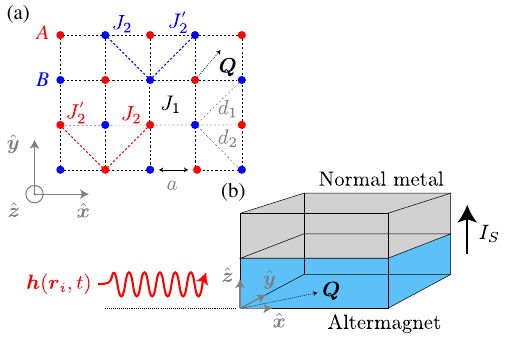}
    \caption{Schematic overview of the altermagnetic insulator – normal metal bilayer setup. (a) The effective, symmetry-determined altermagnetic insulator (AM) Hamiltonian contains three exchange coefficients in addition to an easy-axis anisotropy coupling $K$. The regular Heisenberg coupling $J_1>0$ favours an antiferromagnetic coupling between the diferent sublattices (red and blue) while the intra-sublattice coefficients $J_2<0$ and $J_2'<0$ favours a spatially anisotropic tendency towards ferromagnetic alignment within a sublattice. For the two sublattices, $J_2$, $J_2'$ act along opposite diagonals $d_1$, $d_2$. \textit{a} is the lattice lattice parameter. $\bo{Q}$ is the modulation vector determining the spatial modulation of the magnetic field. It lies generelly in the \textit{xy}-plane and is introduced in Eq. (\ref{eqn: mag}). (b) The bilayer setup showing how a spin current is established across the AM-NM interface by an external magnetic field causing precession of the magnetization in the AM. } 
    \label{fig: grid}
\end{figure}

In this work, we study spin pumping from an altermagnetic insulator into a normal metal using a non-equilibrium Keldysh Green's function perturbation technique. The altermagnet is described using the model of Ref. \cite{cui_arxiv_23}, featuring two intercalated square sublattices with a spin order that breaks $PT$ symmetry. We find that the spin current pumped from the altermagnet is equal to the spin current pumped from a conventional square lattice antiferromagnet with N{\'e}el order when the magnetic field is homogeneous. This can change if the magnetic field used to set the altermagnetic spins into precessional motion is spatially inhomogeneous. Depending on the orientation of the spatial modulation of the magnetic field, the field can couple to degenerate or non-degenerate magnons in the altermagnet, giving rise to novel spin pumping behaviour in the THz regime. Our findings predict that it should be possible to obtain a unique spin pumping signature from a altermagnet/normal metal system which distinguishes it from a regular system based on an antiferromagnet.

\section{Model}

We consider a system consisting of an altermagnetic insulator (AM) coupled to a normal metal (NM) through an interface, described by an exchange term in the Hamiltonian. The setup is presented schematically in Fig. \ref{fig: grid}. The Hamiltonian in question is given by
\begin{equation}
    H=H_\text{AM} + H_\text{N} + H_\text{int}
\end{equation}
where the altermagnetic insulator term is given by the effective, symmetry-determined Hamiltonian
\begin{equation}
    \begin{split}
        H_\text{AM}=&J_1 \sum_{\langle i,j\rangle} \hat{\boldsymbol{S}}_{Ai}\cdot\hat{\boldsymbol{S}}_{Bj} \\
        +&J_2\sum_{\langle i, j \rangle\in{d_1}} \hat{\boldsymbol{S}}_{Ai}\cdot \hat{\boldsymbol{S}}_{Aj}+J_2'\sum_{\langle i, j \rangle\in{d_2}} \hat{\boldsymbol{S}}_{Ai}\cdot\hat{\boldsymbol{S}}_{Aj}+K\sum_i\big(\hat{S}_{Ai}^z \big)^2 \\
        +&J_2'\sum_{\langle i, j \rangle\in{d_1}} \hat{\boldsymbol{S}}_{Bi}\cdot \hat{\boldsymbol{S}}_{Bj}+J_2\sum_{\langle i, j \rangle\in{d_2}} \hat{\boldsymbol{S}}_{Bi}\cdot\hat{\boldsymbol{S}}_{Bj}+K\sum_i\big(\hat{S}_{Bi}^z \big)^2\\
        &-\zeta\sum_{i}\hat{\boldsymbol{S}}_i \cdot \boldsymbol{h}(\bo{r}_i, t) \label{eqn: altermagnetic Hamiltonian}
    \end{split}
\end{equation}
Here, \textit{
A} and \textit{B} denote the two distinct sublattices of the square lattice in the altermagnet and the spin operator $\hat{\boldsymbol{S}}_{Ai}$ corresponds to a localized spin on site \textit{i} residing in the \textit{A} sublattice. As for the various exchange coefficients, $J_1 > 0$ describes the regular Heisenberg exchange between nearest neighbour spins on opposite sublattices, favouring an anti-parallell spin configuration of the sublattices. $J_2 < 0$ and $J_2'<0$ governs a ferromagnetic and spatially anisotropic intra-sublattice exchange coupling, giving rise to a ferromagnetic alignment tendency between sites on the same sublattice, but with unequal strength along the two diagonals $d_1$ and $d_2$. The notation ${\langle i,j\rangle}\in d_i$ is meant to signify a sum over nearest neighbours along one diagonal only.  When $J_2 \neq J_2'$, \textit{PT} symmetry is broken in the system, causing our model to exhibit altermagnetic properties. Upon setting $J_2=J_2'$, the \textit{PT} symmetry is reinstated and our model reduces to an antiferromagnetic insulator with ferromagnetic intra-sublattice exchange. The $J_2$/$J_2'$ anisotropy is taken to be opposite for the \textit{A} and \textit{B} sublattice as is evident from the terms with $J_2$, $J_2'$ in Eq. (\ref{eqn: altermagnetic Hamiltonian}). Finally, $K<0$ determines the easy-axis anisotropy strength along \textit{z} while $\zeta$ is the coupling strength to an external, time-dependent magnetic field $\boldsymbol{h}(t)$ which will drive the spin pumping and which can be non-homogeneous in space. We shall assume the magnetic field to take the general form 
\begin{equation}
    \bo{h}(\bo{r}_i, t)=\cos (\bo{Q}\bo{r}_i) \begin{pmatrix}
        h^x(t), & h^y(t), & 0
    \end{pmatrix} \label{eqn: mag}
\end{equation}
where the vector $\bo{Q}$ sets the spatial period of the magnetic field and where we consider a field without a \textit{z}-component to simplify the calculations. By choosing $\bo{Q}=(0, 0)$, we obtain a spatially homogeneous magnetic field.

The normal-metal Hamiltonian is taken to be a simple tight-binding model with nearest neighbor hopping, diagonalized with momentum-space operators, 
\begin{equation}
    H_\text{N} = \sum_{\boldsymbol{k}\in\square, \sigma}\xi_{\boldsymbol{k}}\cc[]{\boldsymbol{k},\sigma}\ca[]{\boldsymbol{k},\sigma} \label{eqn: normal metal}
\end{equation}
\newline
The square lattice dispersion is given by $\xi_{\boldsymbol{k}}=-2t[\cos (k_x a)+\cos (k_y a)] - \mu$ where \textit{t} the hopping parameter, and $\mu$ the chemical potential. The $\boldsymbol{k}$-sum runs over the first Brillouin zone of the square lattice, denoted by $\square$. 
\begin{figure}[h]
    \centering
    \includegraphics{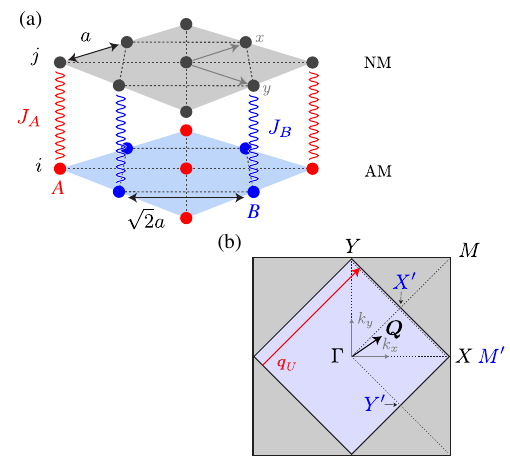}
    \caption{(a) The altermagnetic (AM) and normal metal (NM) layer is coupled by an interfacial, sublattice-dependent exchange coupling. Each site in the normal metal couples to the nearest-neighbour in the altermagnetic layer with the sublattice-dependent $J_A$/$J_B$ exchange coupling strength. (b) The NM Brillouin zone (BZ) is shown in grey while the reduced BZ for momenta defined on the AM sublattices is shown in blue. $q_U=(\pi/a, \pi/a)$ is the Umklapp vector. $\bo{Q}$ is the modulation vector which determines the spatial variation of the magnetic field.}
    \label{fig: exchange coupling}
\end{figure}

The altermagnetic insulator and the normal metal are coupled by an interfacial exchange coupling (see Fig. \ref{fig: exchange coupling}), 
\begin{equation}
    H_\text{int} = -2 \sum_{\langle i,j\rangle | i \in \text{AM}, j\in \text{NM}} J_i \begin{pmatrix}
        \cc[]{j,\uparrow} & \cc[]{j, \downarrow}
    \end{pmatrix}
    \boldsymbol{\sigma} \begin{pmatrix}
        \ca[]{j, \uparrow} \\ \ca[]{j, \downarrow}
    \end{pmatrix}\cdot \hat{\boldsymbol{S}}_i \label{eqn: interfacial exchange interaction}
\end{equation}
\\where $J_i$ for $i\in\{A, B\}$ is the sublattice-dependent interfacial exchange coupling strength and where the sum runs over all sites on the interface. $\boldsymbol{\sigma}$ is the vector of Pauli matrices. The nearest-neighbour notation $\langle i,j \rangle$ indicates that a given site \textit{i} on the interface in the altermagnet couples only to the closest site \textit{j} in the normal metal. 

\subsection{Diagonalizing the altermagnetic Hamiltonian}
Introducing magnon operators for the two sublattices \textit{A} and \textit{B}, the altermagnetic insulator Hamiltonian can, to second order in magnon operators, be written as $H_\text{AM}=\sum_{\boldsymbol{q}\in\lozenge}\psi_{\boldsymbol{q}}^\dagger H_{\boldsymbol{q}} \psi_{\boldsymbol{q}}$ where $\psi_{\boldsymbol{q}}=(\aaa[]{\boldsymbol{q}}, b_{-\boldsymbol{q}}^\dagger)^T$ and 
\begin{equation}
    H_{\boldsymbol{q}}=\begin{pmatrix}
        J^* + 2J_2 \gamma_{1} + 2J_2'\gamma_{2} & 2J_1 \gamma_{3} \\
        2J_1 \gamma_{3} & J^* + 2J_2'\gamma_{1}+2J_2 \gamma_{2}
    \end{pmatrix}
\end{equation}
We have here defined $J^* = J_1 z - (J_2+J_2')z/2 -2K  $ and the structure factors $\gamma_{1}=\cos (q_x a+ q_y a)$, $\gamma_2=\cos (q_x a - q_y a)$   and $\gamma_{3}=\cos q_x a + \cos{q_y a}$. The magnon momenta in the $\boldsymbol{q}$-sum runs over the reduced Brillouin zone (see Fig. \ref{fig: exchange coupling}), denoted by the diamond $\lozenge$, corresponding to the Brilluoin zone of one of the two sublattices. Introducing Bogoliubov quasiparticle operators $\aaaa[]{\boldsymbol{q}}=u_{\boldsymbol{q}}\aaa[]{\boldsymbol{q}}-v_{\boldsymbol{q}}\bbc[]{-\boldsymbol{q}}$, $\ba[]{\boldsymbol{q}}=-v_{\boldsymbol{q}}\aac[]{\boldsymbol{q}}+u_{\boldsymbol{q}}\bba[]{-\boldsymbol{q}}$, 
it follows that for 
\begin{equation}
    u_{\bo{q}}=\frac{1}{\sqrt{2}}\bigg(\frac{1}{\Delta} + 1 \bigg), \qquad v_{\bo{q}}=-\text{sgn}(\gamma_3)\frac{1}{\sqrt{2}}\bigg(\frac{1}{\Delta} - 1\bigg)
\end{equation}
and
\begin{gather}
    \Delta= \sqrt{1-\gamma_e^2}, \\
    \gamma_e=\frac{2J_1\gamma_3}{4J_1 - 2K - (J_2+J_2')(2-\gamma_1-\gamma_2)}
\end{gather}
the coefficients $u_{\bo{q}}$, $v_{\bo{q}}$ diagonalizes the Hamiltonian 
\begin{equation}
    H_\text{AM}=\sum_{\boldsymbol{q}\in\lozenge}\begin{pmatrix}
        \ac[]{\boldsymbol{q}} \\ \ba[]{\boldsymbol{q}}
    \end{pmatrix}
    \begin{bmatrix}
        \omega_{\boldsymbol{q}}^\alpha & 0 \\
        0 & \omega_{\boldsymbol{q}}^\beta
    \end{bmatrix}
    \begin{pmatrix}
        \aaaa[]{\boldsymbol{q}} \\
        \bc[]{\boldsymbol{q}}
    \end{pmatrix} + V
\end{equation}
with the quasiparticle eigenvalues 
\begin{subequations}
    \begin{equation}
        \omega_{\boldsymbol{q}}^\alpha = S\bigg[(J_2 - J_2')(\gamma_{1}-\gamma_{2}) +\frac{2J_1\gamma_3\Delta}{\gamma_e} \bigg] \label{eqn: omega A}
    \end{equation}
    \begin{equation}
        \omega_{\boldsymbol{q}}^\beta = S\bigg[(J_2' - J_2)(\gamma_{1}-\gamma_{2}) +\frac{2J_1\gamma_3\Delta}{\gamma_e} \bigg] \label{eqn: omega B}
    \end{equation}
    \end{subequations}

We note that due to the spatially anisotropic nature of $H_\text{AM}$, a momentum-dependent splitting arises between the two types of magnons in our system,
\begin{align}
    \Delta \omega &= |\omega_{\boldsymbol{q}}^\alpha -\omega_{\boldsymbol{q}}^{\beta}| \\
    &= |2S(J_2 - J_2')(\gamma_{1}-\gamma_{2})| \label{eqn: omega split}
\end{align}
which is generally finite as long as $\bo{q}$ is not located on the high-symmetry paths $\Gamma-M'$ in the reduced Brillouin zone (see Fig. \ref{fig: exchange coupling}).

The interaction term \textit{V} in the diagonalized Hamiltonian stems from the coupling between $\hat{S}^x$/$\hat{S}^y$ and the external magnetic field and is given by 
\begin{multline}
    V=\\-\sum_{\pm}\lambda_{\pm\bo{Q}}\bigg[\big(\alpha_{\pm\bo{Q}}^{\vphantom{\dagger}} + \beta_{\pm\bo{Q}}^\dagger \big)h^-(t) + \big(\alpha_{\pm\bo{Q}}^{{\dagger}} + \beta_{\pm\bo{Q}}^{\vphantom{\dagger}} \big)h^+(t) \bigg] \label{eqn: interaction}
\end{multline}
where we have defined $\lambda_{\pm\bo{Q}}=\zeta\frac{\sqrt{N_A S}
}{4}(u_{\pm\bo{Q}}+v_{\pm\bo{Q}})$, $h^\pm=h_x\pm i h_y$ and where $N_A=N/2$ is the number of lattice sites in the \textit{A} sublattice. We shall assume throughout this paper the number of lattice sites in the \textit{A} and \textit{B} sublattice to be equal. Recall that $\bo{Q}$ characterizes the spatial inhomogeneity of the external magnetic field. 
\subsection{Interfacial exchange interaction}

We now proceed by considering the interfacial exchange interaction between the altermagnetic insulator and the normal metal. Introducing magnon operators and performing a Bogoliubov transformation, Eq. (\ref{eqn: interfacial exchange interaction}) can be written as 
\begin{equation}
    H_\text{int}=H_\text{int}^\parallel + H_\text{int}^z 
\end{equation}
where we to first order in magnon operators have defined (see appx. \ref{sec:appA} for details)
\begin{widetext}
    \begin{align}
        H_\text{int}^\parallel &= -\sum_{\boldsymbol{q}\in\lozenge}\sum_{\boldsymbol{k}\in \square}\sum_{\kappa \in \{R,U\}}\sum_{\nu \in \{\alpha, \beta^\dagger \}}M_{\boldsymbol{q}}^{\nu \kappa}\nu_{\boldsymbol{q}}\cc[]{\boldsymbol{k}^\kappa, \downarrow}\ca[]{\boldsymbol{k}-\boldsymbol{q},\uparrow} + \text{h.c.} \label{eqn: interaction parallel}\\
        H_\text{int}^z &= -\sqrt{2S}\sum_{\boldsymbol{k}\in \square}\sum_{\kappa\in\{R,U\}}(\overline{J}_A - \kappa\overline{J}_B)\big(\cc[]{\boldsymbol{k}^\kappa, \uparrow}\ca[]{\boldsymbol{k},\uparrow}-\cc[]{\boldsymbol{k}^\kappa, \downarrow}\ca[]{\boldsymbol{k}, \downarrow}\big) \label{eqn: interaction z}
    \end{align} 
\end{widetext}
Here we have defined $\kappa=R=1$ and $\boldsymbol{k}^\kappa=\boldsymbol{k}^R=\boldsymbol{k}$ for the regular scattering process and $\kappa=U=-1$ and $\boldsymbol{k}^\kappa=\boldsymbol{k}^U=\boldsymbol{k}+\boldsymbol{q}_U$ for the Umklapp process where the NM momenta $\boldsymbol{k}$ falls outside the reduced Brillouin zone of the altermagnetic insulator. $\boldsymbol{q}_U=(\pi/a, \pi /a)$ is the Umklapp vector connecting the momenta $\boldsymbol{q}$ in the reduced Brillouin zone of the sublattices with the regular square lattice momenta $\boldsymbol{k}$. The coefficients $M_{\boldsymbol{q}}^{\nu \kappa}$ in Eq. (\ref{eqn: interaction parallel}) are defined as
\begin{subequations}
\begin{align}
    M_{\boldsymbol{q}}^{\alpha \kappa} &= \overline{J}_A u_{\boldsymbol{q}} + \kappa\overline{J}_B v_{\boldsymbol{q}} \\
    M_{\boldsymbol{q}}^{\beta^\dagger \kappa} &= \overline{J}_A v_{\boldsymbol{q}} + \kappa\overline{J}_B u_{\boldsymbol{q}}
\end{align}
\end{subequations}
and the modified exchange coefficients $\overline{J}_A$/$\overline{J}
_B$ are defined as 
\begin{equation}
    \overline{J}_A = 2\frac{\sqrt{2S}J_A N_A}{N_N \sqrt{N_A}}, \qquad \overline{J}_B = 2\frac{\sqrt{2S}J_B N_B}{N_N \sqrt{N_B}}
\end{equation}
Here, $N_N$ is the number of normal metal sites at the interface while $N_A (N_B)$ is the number of sites on the \textit{A} (\textit{B}) sublattice at the interface. For the geometry studied in this paper, the number of sites at the interface is identical to the number of sites in general due to the 2D geometry.

\section{Spin current}
In order to obtain the spin current caused by the time-dependent external magnetic field in the altermagnet, we follow a similar approach as laid out by Kato \etal \cite{kato_prb_19}. We consider the time-dependence of the magnetization in the normal metal
\begin{equation}
    I_S = -\frac{d}{dt}\langle s^z \rangle = -i[H, s^z]
\end{equation}
where all operators above are in the Heisenberg picture and the normal metal magnetization is given by 
\begin{equation}
s^z=\frac{1}{2}\sum_{\boldsymbol{k}\in\square} (\cc[]{\boldsymbol{k},\uparrow}\ca[]{\boldsymbol{k},\uparrow} - \cc[]{\boldsymbol{k},\downarrow}\ca[]{\boldsymbol{k},\downarrow})
\end{equation} 
As the fermionic number operators in the normal metal commutes with the bosonic magnon operators of the altermagnetic insulator as well as with the fermionic number operators of $H_\text{int}^z$,  $[H_\text{AM}, s^z] = [H_{\text{int}}^z, s^z] = 0$, and the only non-zero commutator to consider is the one with the in-plane interaction term $H_\text{int}^\parallel$. Writing out the time-dependence and denoting the Heisenberg picture fermion and magnon operators via the superscript $H$, we get:
\begin{widetext}
\begin{equation}
\begin{split}
    [H_\text{int}^\parallel, s^z]&=-\sum_{\boldsymbol{q}\in\lozenge}\sum_{\boldsymbol{k}\in\square}\sum_{\kappa\in\{R, U\}}\sum_{\nu\in\{\alpha, \beta^\dagger\}}M_{\boldsymbol{q}}^{\nu \kappa}\nu_{\boldsymbol{q}}^{{H}}{(t)}\cc[H]{\boldsymbol{k}^\kappa, \downarrow}{(t)}\ca[{{H}}]{\boldsymbol{k}-\boldsymbol{q},\uparrow}{(t)} - \text{h.c.} \\
    \Rightarrow I_S{(t)} &= \text{Re}\bigg\{-2i \sum_{\boldsymbol{q}\in\lozenge}\sum_{\kappa\in\{R, U\}}\sum_{\nu\in\{\alpha, \beta^\dagger\}} \big\langle M_{\boldsymbol{q}}^{\nu\kappa} \nu_{\boldsymbol{q}}^{{H}}{(t)}s_{\boldsymbol{q}}^{\kappa-,H}{(t)} \big\rangle \bigg\} \\
    &=2\sum_{\boldsymbol{q}\in\lozenge}\sum_{\kappa\in\{R, U\}}\sum_{\nu\in\{\alpha, \beta^\dagger\}}\text{Re} \{G_{\boldsymbol{q},\kappa,\nu}^< (t,t)\}
    \label{eqn: spin current commutator}
    \end{split}
    \end{equation}
\end{widetext}
where $I_S(t) \equiv \langle I_S \rangle$.
Here we have introduced the operator 
\begin{equation}
    s_{\boldsymbol{q}}^{\kappa-,H}(t)\coloneqq \sum_{\boldsymbol{k}\in\square} \cc[{H}]{\boldsymbol{k}^\kappa, \downarrow}(t)\ca[{H}]{\boldsymbol{k}-\boldsymbol{q}, \uparrow}(t) \label{eqn: s_operator}
\end{equation}
and defined the lesser Green's function 
\begin{equation}
G_{\boldsymbol{q},\kappa,\nu}^<(t_1,t_2) = -i\big\langle M_{\boldsymbol{q}}^{\nu\kappa} \nu_{\boldsymbol{q}}^{{H}}(t_1)s_{\boldsymbol{q}}^{\kappa-,{H}}(t_2) \big\rangle \label{eqn: lesser greens function}
\end{equation}
The average $\langle ... \rangle$ is taken in the ground-state of the full time-dependent Hamiltonian. The terminology ``lesser Green's function" is here used for this expectation value only because it will be evaluated by applying Langreth rules \cite{Rammer_revmodphys_rmp_86} on a corresponding contour-ordered expectation value for a choice of $t_1$ and $t_2$ that normally correspond to a lesser Green's function.

In order to arrive at a final expression for the spin current $I_S$, we need to obtain an expression for the lesser Green's function in Eq. (\ref{eqn: lesser greens function}) at equal times, $G_{\boldsymbol{q},\kappa,\nu}^<(t,t)$, including any relevant corrections due to the presence of a time-dependent magnetic field and the interfacial exchange interaction. This can be done by considering the related, contour-ordered Green's function 
\begin{equation}
    G_\mathcal{C}(t_1, t_2) = -i \langle T_\mathcal{C}  M_{\boldsymbol{q}}^{\nu\kappa}[ \nu_{\boldsymbol{q}}^{H}(t_1)s_{\boldsymbol{q}}^{\kappa-,{H}}(t_2) ]\rangle
\end{equation}
where $T_\mathcal{C}$ is the contour-ordering operator and where the subscript $\mathcal{C}$ of the Green's function indicates that the time parameters $t_1$, $t_2$ are defined on the Keldysh contour. We will explicitly choose $t_1$ to reside on the forward path and $t_2$ on the backward path. For this choice of time-parameters, it follows that the contour-ordered Green's function equals its lesser component, 
\begin{equation}
    G_\mathcal{C}(t_1, t_2)=G_\mathcal{C}^<(t_1, t_2)=-i\langle  M_{\boldsymbol{q}}^{\nu\kappa}\nu_{\boldsymbol{q}}^{H}(t_1)s_{\boldsymbol{q}}^{\kappa-,{H}}(t_2)\rangle,
\end{equation} \\
which is precisely the expectation value we required to compute the spin current $I_S(t)$.
\newline

To proceed, we will compute the lesser component $G_\mathcal{C}^<$ in the interaction picture, considering the exchange interaction at the interface [Eqs. (\ref{eqn: interaction parallel})-(\ref{eqn: interaction z})] as a perturbation in the Keldysh formalism. The formal perturbation expansion is now given by \cite{Rammer_revmodphys, stefanucci2013nonequilibrium} 

\begin{multline}
    G_\mathcal{C}(t_1, t_2)=-iM_{\boldsymbol{q}}^{\nu\kappa}\big\langle T_\mathcal{C}  \nu_{\boldsymbol{q}}(t_1)s_{\boldsymbol{q}}^{\kappa,-}(t_2)\\ \times e^{-i\int_\mathcal{C}dt H_\text{int}(t)} \big\rangle _0\label{eqn: formal expansion} 
\end{multline}
 The average $\langle ... \rangle_0$ is now taken in the absence of the interface interaction $H_\text{int}$ and the operators are defined in the interaction picture. We omit a superscript $I$ on the operators for brevity of notation. In the following, we can ignore the contribution to $H_\text{int}$ from $H_\text{int}^z$ as the respective expectation values caused by this term will always be odd in magnon operators. We thus proceed by considering the parallel term $H_\text{int}^\parallel$ only. Expanding Eq. (\ref{eqn: formal expansion}) to first order in the perturbation and rewriting the contour-ordered Green's function as a Green's function over the regular time axis via the Langreth rules \cite{stefanucci2013nonequilibrium}, we obtain the following, lowest-order, non-vanishing contribution to the lesser Green's function in Eq. (\ref{eqn: spin current commutator}) (see Appx. \ref{sec:appB} for details) 
\begin{widetext}
    \begin{equation}
    G_{\boldsymbol{q},\kappa,\nu}^<(t,t)=-i\int\frac{d\omega}{2\pi}\sum_{\kappa'\in\{R,U\}}M_{\boldsymbol{q}}^{\nu\kappa}(M_{\boldsymbol{q}}^{\nu\kappa'})^*\big[G_{\nu, \boldsymbol{q}}^R(\omega)G_{s^+,\kappa\kappa', \boldsymbol{q}}^{ <}(\omega) + G_{\nu, \boldsymbol{q}}^<(\omega)G_{s^+,\kappa\kappa', \boldsymbol{q}}^{ A} (\boldsymbol{q},\omega) \big]\label{eqn: lesser component important}
\end{equation}
\end{widetext}
where we have introduced the operator $s_{\boldsymbol{q}}^{\kappa,+}=(s_{\boldsymbol{q}}^{\kappa,-})^\dagger$. If we let $\varphi \in \{ \alpha, \beta^\dag, s^+\}$, the corresponding lesser, retarded and advanced Green's functions appearing in the equation above are given by
\begin{subequations}
    \begin{align}
        G_{\varphi,\boldsymbol{k}}^< (t_1, t_2) &= -i \langle \varphi_{\boldsymbol{k}}^\dagger(t_2)\varphi_{\boldsymbol{k}}^{\vphantom{\dagger}}(t_1) \rangle _0 \\
        G_{\varphi,\boldsymbol{k}}^R (t_1, t_2) &= -i\theta(t_1-t_2) \langle [\varphi_{\boldsymbol{k}}^{\vphantom{\dagger}}(t_1),\varphi_{\boldsymbol{k}}^{\dagger}(t_2)] \rangle _0 \\
        G_{\varphi, \boldsymbol{k}}^A (t_1, t_2) &= i\theta(t_2-t_1) \langle [\varphi_{\boldsymbol{k}}^{\vphantom{\dagger}}(t_1),\varphi_{\boldsymbol{k}}^{\dagger}(t_2)]  \rangle _0
    \end{align}
\end{subequations}
where the 0 subscript indicates an expectation value in the interaction picture, taken in the absence of the time-dependent perturbations, thus making it equivalent to the Heisenberg picture without the interaction term $V$.

We now use that the advanced and retarded Green's function component is related by $G_{\varphi, \boldsymbol{k}}^R(\omega)=[G_{\varphi,\boldsymbol{k}}^A(\omega)]^*$ as well as the definition of the distribution function, 
\begin{equation}
    f^\varphi (\omega, \boldsymbol{k}) = \frac{G_{\varphi,\boldsymbol{k}}^< (\omega)}{2i \text{Im}\{G_{\varphi, \boldsymbol{k}}^R (\omega) \}} \label{eqn: distribution function}
\end{equation}
which in equilibrium is equal to the Bose-Einstein distribution $n_B(\omega, T)$ (for bosons), and also for the composite boson operator $s^+$ in the normal metal which is bilinear in fermion operators. This can be proven using the Lehmann-representation for the Green's function in equilibrium. Out-of-equilibrium, Eq. (\ref{eqn: distribution function}) 
also holds as the definition of the distribution function. This can be seen by noting that the particle density $n = \langle \varphi^\dag(\vecr,t) \varphi(\vecr,t)\rangle$ for particle type $\varphi$ is proportional to the lesser Green's function $G_\varphi^<$. At the same time, $n$ should also be determined by an integral over the spectral weight times the distribution function for particle $\varphi$. Put differently, the number of particles should equal an integral over the available states times the probability that they are occupied. Since the denominator of Eq. (\ref{eqn: distribution function}) is precisely the spectral weight, moving it over to the left hand side then shows how the particle density (determined by an integral over the lesser Green's function) is equal to an integral over the spectral weight times $f^\varphi$. Thus, $f^\varphi$ is the distribution function also out-of-equilibrium.

\section{Corrections to Green's functions}
In order to evaluate the spin current expression in Eq. (\ref{eqn: spin current commutator}) through the equal-time lesser component in Eq. (\ref{eqn: lesser component important}), we will need Green's functions corresponding to the operators $\alpha$, $\beta^\dagger$ and $s^+$. Below, we compute these quantities.

\subsection{Spin pumping in the altermagnet}

We consider first spin pumping in the altermagnet and the subsequent corrections to the $\alpha$/$\beta^\dagger$ magnon Green's functions. We consider the interaction given by $V$ (Eq. (\ref{eqn: interaction})) as a perturbation in the interaction picture. It can be shown that to second order in the interaction (the reader is referred to Appx. \ref{sec:appC} for details), the retarded components of the Green's functions remain unaltered,
\begin{subequations}
\begin{align}
    G_{\alpha,\boldsymbol{q}}^R(\omega)=\frac{1}{\omega-\omega_{\boldsymbol{q}}^\alpha + i\eta^\alpha} \\
    G_{\beta^\dagger, \boldsymbol{q}}^R(\omega)=-G_{\beta,\boldsymbol{k}}^{A}(-\omega)=\frac{1}{\omega + \omega_{\boldsymbol{q}}^{\beta} +i\eta^{\beta}}
\end{align}
\end{subequations}
where $\eta^\nu$ can be considered the inverse lifetimes of the $\alpha$, $\beta$ magnons. The lesser component of the Green's functions picks up a second-order correction on the form 
\begin{multline}
    \Delta G_{\nu,\bo{q}}^< (\omega)=-4\pi i h_0^2 \delta(\omega-\Omega) \\\times\sum_{\pm}\lambda_{\pm\bo{Q}}^2 |G_{\nu, \pm\bo{Q}}^R(\omega)|^2 \delta_{\bo{q}, \pm\bo{Q}} \label{eqn: lesser correction}
\end{multline}
where $\nu\in\{\alpha,\beta^\dagger\}$ and $\lambda_{\pm\bo{Q}}$ is defined under Eq. (\ref{eqn: interaction}). Using Eq. (\ref{eqn: distribution function}), it then follows that the $\alpha$/$\beta^\dagger$ magnon distribution functions picks up a correction due to the spin pumping, given to second order in the perturbation as 
\begin{equation}
    f^\nu (\omega, \boldsymbol{k})=n_B(\omega, T) + \frac{2\pi h_0^2}{\eta^\nu}\delta(\omega-\Omega)\sum_{\pm }\lambda_{\pm\bo{Q}}^2 \delta_{\bo{q}, \pm\bo{Q}} \label{eqn: magnon distribution correction}
\end{equation}
\subsection{Normal-metal spin susceptibility}
We proceed by considering the retarded component of the spin susceptibility $s^+$ operator introduced in Eq. (\ref{eqn: s_operator}). Utilizing that the normal-metal Hamiltonian is diagonal in momentum indices, it is straight-forward to obtain the imaginary-time Matsubara Green's function $\overline{G}_{s^+,\boldsymbol{q}}(i\omega_n)$. The retarded susceptibility Green's function is then obtained by an analytical continuation 
\begin{equation}
    G_{s^+, \boldsymbol{q}}^R(\omega)=\overline{G}_{s^+,\boldsymbol{q}}( i\omega_n \rightarrow \omega+i\eta)
\end{equation}

The retarded component is given by (the reader is referred to Appx. \ref{sec:appD} for details)
\begin{equation}
    G_{s^+,\kappa\kappa',\boldsymbol{q}}^R (\omega)=\delta_{\kappa,\kappa'}\sum_{\boldsymbol{k}\in\square}\frac{n_F(\xi_{\boldsymbol{k}-\boldsymbol{q}}, T)-n_F(\xi_{\boldsymbol{k}^\kappa}, T)}{\omega + \xi_{\boldsymbol{k}-\boldsymbol{q}}-\xi_{\boldsymbol{k}^\kappa} + i\eta^N}
\end{equation}
\\where $n_F$ is the Fermi-Dirac distribition.

We now have all ingredients necessary to construct an explicit expression for the spin current. We take into account the appropriate corrections to the Green's functions due to spin pumping in the altermagnet as well as the perturbation caused by the interfacial exchange interaction. We also consider the relation between lesser components and the distribution function given in Eq. (\ref{eqn: distribution function}). It then follows that the spin current from Eq. (\ref{eqn: spin current commutator}) becomes 
\begin{widetext}
    \begin{align}
    I_s(t)&=-\sum_{\bo{q}\in\lozenge}\sum_{\kappa\in\{R,U\}}\sum_{\nu\in\{\alpha,\beta^\dagger\}}\text{Re}\{G_{\bo{q}, \kappa, \nu}^<(t,t)  \} \\
    &=-4\pi h_0^2\int \frac{d\omega}{2\pi}\delta(\omega-\Omega)\sum_{\bo{q}\in \lozenge}\sum_{\kappa\in\{R,U\}}\sum_{\nu\in\{\alpha,\beta^\dagger\}}|M_{\bo{q}}^{\nu\kappa}|^2\frac{1}{\eta^\nu}\sum_{\pm}\lambda_{\pm\bo{Q}}^2\delta_{\bo{q},\pm\bo{Q}}\text{Im}\big\{G_{\nu, \bo{q}}^R(\omega) \big\}\text{Im}\big\{G_{s^+, \kappa, \bo{q}}^R(\omega) \big\} \\ 
    &=-2 h_0^2 \sum_{\kappa\in\{R,U\}}\sum_{\nu\in \{\alpha, \beta^\dagger\}}\sum_{\pm}|M_{\pm\bo{Q}}^{\nu\kappa}|^2\frac{\lambda_{\pm\bo{Q}}^2}{\eta^\nu}\text{Im}\big\{G_{\nu, \pm \bo{Q}}^R(\Omega) \big\}\text{Im}\big\{G_{s^+, \kappa, \pm\bo{Q}}^R(\Omega) \big\} \label{eqn: spin_current_green}
\end{align}
and using the explicit forms of the Green's functions obtained above, we arrive at the final expression for the spin current,
\begin{multline}
    I_s=\frac{NS\zeta^2 h_0^2}{8}\sum_{\kappa\in\{R,U\}}\sum_{\pm}(u_{\pm\bo{Q}} + v_{\pm\bo{Q}})^2\bigg(\frac{|M_{\pm\bo{Q}}^{\alpha\kappa}|^2}{(\hbar\Omega - \omega_{\pm\bo{Q}}^\alpha)^2 + (\eta^\alpha)^2 } \\+ \frac{|M_{\pm\bo{Q}}^{\beta^\dagger\kappa}|^2}{(\hbar\Omega + \omega_{\pm\bo{Q}}^{\beta})^2 + (\eta^{\beta})^2 } \bigg) \sum_{\bo{k}\in\square}\text{Im}\bigg\{\frac{n_F(\xi_{\bo{k}-\bo{Q}}, T) - n_F(\xi_{\bo{k}^\kappa}, T)}{\hbar\Omega + \xi_{\bo{k}-\bo{Q}}-\xi_{\bo{k}^\kappa}+i\eta^N}\bigg\} \label{eqn: spin-current final}
\end{multline}
\end{widetext}
Here we have used that the spin pumping in the altermagnet does not affect the normal-metal Hamiltonian, so that $f^{s+,\kappa}(\omega, \boldsymbol{k})=n_B(\omega, T)$ where $T$ is temperature. 

\section{Results and discussion}
The final expression for the spin current lends itself to a simple interpretation. To first order in the external magnetic field, the field couples to magnons with wavevector $\pm\bo{Q}$ in the altermagnet. The result of this is that the spin current is peaked around the resonance frequencies $\hbar\Omega_R=  \omega_{\pm\bo{Q}}^\alpha$ and $\hbar\Omega_R=-  \omega_{\pm\bo{Q}}^\beta$. This entails that the frequency at which the resonance takes place depends directly on the various exchange coefficients $J_1$, $J_2$ and $J_2'$ in addition to the anisotropy $K$.  The magnitude of the spin current response is affected by the imaginary part of the Lindhard function for the given frequency $\Omega$. If one assumes the NM broadening $\eta^N$ to be small, we can approximate the imaginary part of the Lindhard function as 
\begin{multline}
    \text{Im}\big\{G_{s^+, \kappa, \pm\bo{Q}}^R(\Omega) \big\}\\\simeq\sum_{\bo{k}\in\square}\big[n_F(\xi_{\bo{k}\mp\bo{Q}}) - n_F({\xi}_{\bo{k}^\kappa})\big]\delta(\hbar\Omega + \xi_{\bo{k}\mp\bo{Q}}-\xi_{\bo{k}^\kappa}) \label{eqn: Lindhard delta}
\end{multline}
 For a given choice of $\bo{Q}$ (the spatial modulation of the external field), this is simply put a sum over all physically available transitions in the NM separated by the energy $\hbar \Omega$. It is thus clear that depending on the frequency $\Omega$, the magnitude of the spin current will be affected by the availability of electron states in the NM that have transitions between them due to the magnons reflecting off the interface and imparting energy, spin, and momentum to the normal metal. Specifically, the spin pumping field excites magnons in the altermagnet and these cause spin-flip transitions in the normal metal which results in a net spin transfer across the interface. These spin-flip events must be such that $\Omega$ matches an allowed transition between electron states that conserves energy and momentum up to a reciprocal lattice vector in the altermagnet, the latter enabling Umklapp scattering.  

Before moving on, we briefly mention that we expect the presence of spin-flip scattering in the NM, for instance mediated by magnetic impurities, to increase the magnitude of the spin pumping current. In the absence of spin-flip scattering, the spin pumping in the AM will lead to a spin accumulation in the NM which will counteract the spin current. If we instead have considerable spin-flip scattering in the NM, the accumulation will be prevented by spin conversion causing the NM to behave as an effective spin sink. With this consideration in mind, we also expect that upon increasing the thickness of a NM with spin-flip scattering, the magnitude of the spin current will increase. 

We now proceed by considering the spin current for different external field configurations.

\subsection{Spin current for a homogeneous magnetic field}
A homogeneous magnetic field is modelled by setting $\bo{Q}=(0.0,0.0)$. Due to the form of the altermagnetic magnon correction caused by the external magnetic field, the spin current expression simplifies significantly in this case. The key observation is that in this case, the spin pumping correction to the magnon distribution functions $f^\nu(\omega, \boldsymbol{k})$ given in Eq. (\ref{eqn: magnon distribution correction}) is restricted to the spatially homogeneous mode $\boldsymbol{q}=(0.0,0.0)$ through the Kronecker delta $\delta_{\boldsymbol{q},\pm\boldsymbol{Q}}$ from Eq. (\ref{eqn: lesser correction}). This can be intuitively be understood as a consequence of the external magnetic field $\boldsymbol{h}$ carrying no spatial dependence, thus causing a coupling between the external field and the uniform magnon modes only, to the chosen order of magnon operators. While a consideration of the uniform mode in spin pumping is not unusual in itself \cite{cheng_prl_14, Kamra_spinpumping}, the effect in this context is that the altermagnetic nature of the AM lattice disappears in the spin pumping contribution, leaving only the regular antiferromagnetic contribution. 

The disappearance of the altermagnetic character can be understood by considering the magnon dispersions in Eq. (\ref{eqn: omega A})-(\ref{eqn: omega B}) as well as the level splitting between the $\alpha$ and $\beta$ magnons, given in Eq. (\ref{eqn: omega split}). While these for general momenta $\boldsymbol{q}$ show a lifted degeneracy between the $\alpha$/$\beta$ branches, the splitting vanishes in the $\boldsymbol{q}\rightarrow \Gamma$ limit, as well as when $\bo{q}$ lies on the $\Gamma-M'$ line (consider the magnon dispersion in Fig. \ref{fig: magnon dispersion}). In particular at $\bo{q}=\Gamma$, the altermagnetic character of the system vanishes  completely for the square lattice with coordination number $z=4$ no matter the values of $J_2$, $J_2'$. When we then take into account that to first order in the perturbation, the external field with wave vector $\bo{Q}$ couples directly to magnons with wave vector $\bo{q}=\pm\bo{Q}$, the problem becomes obvious. In the present model and for a homogeneous field, the alter- and antiferromagnetic insulator is identical in terms of using the magnons in the magnetic insulator to drive a spin current. 

\subsection{Spin current for a non-homogeneous magnetic field}
\begin{figure}
    \centering
    \includegraphics{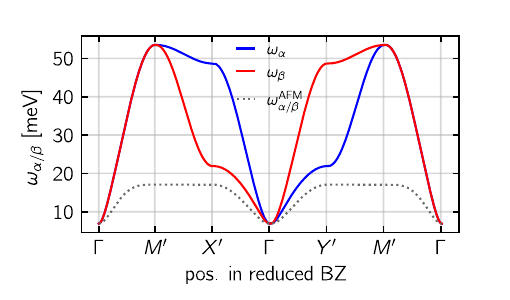}
    \caption{Magnon spectra are shown for $\alpha$ (blue) and $\beta$ (red) magnons in an altermagnet with $J_1=3.90$, $J_2=-7.90$, $J_2'=-1.21$ and $K=-0.73 \text{ meV}$. The dotted line shows the degenerate AFM magnon spectrum for the same $J_1$, but with $J_2=J_2'=0.0$. We observe that the three branches are all the same at the $\Gamma$ point. Additionally we note that the $\Gamma-M'$ is a special path in the BZ where the AM magnon spectrum is degenerat, in general the splitting is non-zero away from $\Gamma$.}
    \label{fig: magnon dispersion}
\end{figure}

\begin{figure}[ht]
    \centering
    \includegraphics[width=\linewidth]{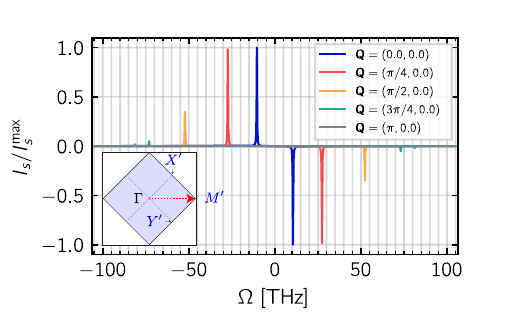}
    \caption{Altermagnet-normal metal spin current $I_s$ as function of field frequency $\Omega$, normalized on the maximum AFM spin current with a uniform magnetic field using $J_1=3.90$, $J_2=-7.90$, $J_2'=-1.21$ and $K=-0.73 \text{ meV}$. The $\bo{Q}=(0.0,0.0)$ spin current is essentially the AFM-NM spin current as AM character vanishes at the $\Gamma$ point. We let $\bo{Q}$ move along $\Gamma-M'$, indicated by the red arrow in the inset, which is the $X$ point in the NM Brillouin zone. Along this path, the AM magnon spectrum is degenerate. This entails that the resonance frequency $\Omega_R$ changes with $\bo{Q}$, but that we have the same resonance frequencies for positive and negative $\Omega$. As $\text{Im}\{G_{s^+}(\Omega) \} = -\text{Im}\{G_{s^+}(-\Omega) \} $, the postive and negative frequency peaks share the same magnitude, but with opposite sign.  }
    \label{fig: AM along x}
\end{figure}

Due to the above observations, we proceed directly to the consideration of a non-homogeneous external field comparing it to the homogeneous $\bo{Q}=(0.0, 0.0)$ AFM spin current. We use a NM hopping parameter $t=500 \text{ meV}$, $J_1=3.90 \text{ meV}$, $J_2=-7.90  \text{ meV}$, $J_2'=-1.21\text{ meV}$ as well as an easy-axis anisotropy $K=-0.73 \text{ meV}$. The particular exchange coefficients are obtained through \textit{ab initio} methods and originally presented in Ref. \cite{cui_arxiv_23}. While the results in the following depend on these coefficients, our findings are general and expected to be generally present for systems with the appropriate altermagnetic symmetries. Finally, the chemical potential in the NM is set to $\mu=0.0$, the half-filled case. We set the magnon broadening to $\eta^{\alpha/\beta}=0.1\text{ meV}$ and the NM broadening, modelling \eg inelastic electron-phonon or electron-electron interactions, to $\eta^N=1\text{ meV}$ and use a recipropcal temperature $1/k_B T=0.1 \text{ meV}^{-1}$, corresponding to a temperature of around $116$ K.

\begin{figure}[ht]
    \centering
    \includegraphics[width=\linewidth]{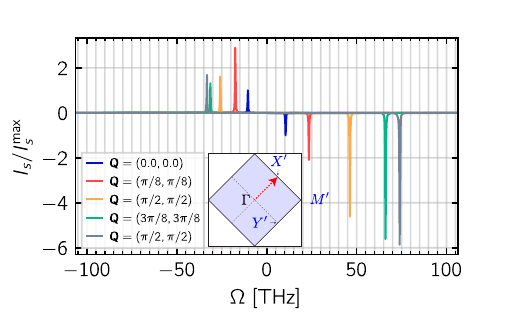}
    \caption{Altermagnet-normal metal spin current $I_s$ as function of field frequency $\Omega$, normalized on the maximum AFM spin current with a uniform magnetic field using $J_1=3.90$, $J_2=-7.90$, $J_2'=-1.21$ and $K=-0.73 \text{ meV}$. We let $\bo{Q}$ move along $\Gamma-X'$, indicated by the red arrow in the inset, which is halfway towards the NM BZ $M$ point. We observe that a consequency of the splitting between $\alpha$ and $\beta$ magnons, the resonance frequencies at which the spin current peaks are located, are no longer symmetric around $\Omega=0$. We also observe that the magnitudes of the positive and negative frequency peak are now different as $\text{Im}\{G_{s^+}(\Omega) \}$ now gives a different weight for the postive and negative resonance frequency.}
    \label{fig: AM along diag}
\end{figure}

We begin by considering an external field whose spatial modulation is oriented along the $\Gamma-M'$ line, depicted in Fig. \ref{fig: AM along x}. Along this high-symmetry path, the magnon branches are degenerate. We observe that as $\bo{Q}$ increases in magnitude, the resonance frequencies $\Omega_R$ increase as well. Considering the magnon dispersion in Fig. \ref{fig: magnon dispersion}, this is as expected as we are coupling higher-energy magnons with increasing $\bo{Q}$. The resonance peaks are symmetric around $\Omega=0$ and equal in magnitude for $\Omega=\pm\Omega_R$ which is as expected since $\text{Im}\{G_{s^+}(\Omega) \} = -\text{Im}\{G_{s^+}(-\Omega) \} $. This behaviour is qualitatively similar to that of spin pumping in a AFM/NM bilayer. As $\bo{Q}$ moves away from the $\Gamma$-point, the energy of the magnons which couple to the field changes with a subsequent shift in the resonance frequency $\Omega_\text{R}$. As the magnon branches are degenerate along this path in the AM Brillouin zone, the resonance frequencies $\Omega_\text{R}$ are symmetric around zero. Recall that the positive and negative resonance frequency is set by $\omega_{\pm\bo{Q}}^\alpha $ and $-\omega_{\pm\bo{Q}}^\beta$ respectively, the magnitudes of which is the same for a degenerate dispersion. For an AFM, this behaviour is expected for any $\bo{Q}$ in the BZ due to the degenerate magnon branches. This suggests that for $\bo{Q}$ lying on the $\Gamma-M'$ path in the AM BZ, the spin pumping behaviour is qualitatively similar to that of an AFM, but with resonance frequencies and magnitudes dependent on the various exchange coefficients as well as the anisotropy. The effect of $\bo{Q}$ on the magnitude of the spin current is discussed at the end of this section.

The similarities between AFM and AM spin pumping vanishes when we consider $\bo{Q}$ away from the $\Gamma-M'$ line of magnon degeneracy, i.e. the most probable case as the degeneracy is present only exactly at the $\Gamma-M'$ line. Then, we couple to $\alpha$/$\beta$ magnons with different energies $\omega^\alpha \neq \omega^\beta$. Taking into consideration that the $\alpha$, $\beta$ magnons give rise to the postive and negative resonance respectively, the breaking of the magnon degeneracy implies the emergence of an asymmetry between the positive and negative resonance frequencies. This is also clear from Fig. \ref{fig: AM along diag} where we consider $\bo{Q}$ on the $\Gamma-X'$ line in the reduced BZ. Along this path, $\alpha$ magnons have higher energies than the $\beta$ magnons, something which manifests itself in the higher magnitudes of the positive resonance frequencies as opposed to the negative ones. We emphasize that this asymmetry is not a fine-tuning effect, but rather the expected behaviour for an arbitrary spatial modulation $\bo{Q}$. It is only when the modulation matches with the high-symmetry paths of magnon degeneracy that we regain the frequency-symmetric characteristics associated with the AFM. We thus expect that this should be possible to verify experimentally. As mentioned above, the magnitude of the spin current is determined in part by the Lindhard function, i.e. by the presence of available transitions in the NM. As the positive and negative resonance now is different, their magnitude given in part by the susceptibility Lindhard function is also in general different, giving rise to a highly asymmetrical spin current behaviour which is both asymmetric in resonance frequency as well as magnitude. 

We finally mention that the magnitudes of the spin currents depicted in Fig. \ref{fig: AM along x} and \ref{fig: AM along diag}, normalized on the $\bo{Q}=(0.0,0.0)$ AFM spin current, is highly dependent on the specific parameters of the system. In general, the energy of the AM magnons is on the order of 1-10 \text{meV} which as is well known gives rise to the desireable $\text{THz}$ frequency response. The energy of the NM electrons, reflected through the hopping parameter \textit{t}, is generally on the order of 1 eV. This mismatch of energy scales will generally decrease the number of available transitions in the NM, shown in Eq. (\ref{eqn: Lindhard delta}). This can give rise to large observed variation in spin current magnitudes due to the interplay between the resonance frequency $\Omega_{R}$ set by the AM magnon dispersion and the particular number of available transitions at the specific value of $\bo{Q}$. We nevertheless draw the conclusion that for $\bo{Q}$ close to the $\Gamma$-point, the results indicate that it is possible to obtain a spin current which is comparable in magnitude to a regular antiferromagnet, but with novel features such as the non-symmetric resonance frequencies and subsequent pumping modified magnitudes. We expect that this also holds when replacing the normal metal with a superconductor, as has recently been studied \cite{fyhn_prb_21} in the antiferromagnetic case.

\section{Summary}
We have computed spin pumping from an altermagnetic insulator into a normal metal using a non-equilibrium Keldysh Green's function perturbation technique. The altermagnetic model introduced in Ref. \cite{cui_arxiv_23}, consisting of two intercalated square sublattices with a spin order that breaks $PT$ symmetry, was used. Our calculations show that for a homogeneous magnetic field, the spin current pumped from the altermagnet is the same as the spin current pumped from a conventional square lattice antiferromagnet with N{\'e}el order.  When the magnetic field becomes spatially dependent, however, the spin pumping characteristics becomes dependent on the altermagnetic crystal orientation. Along the high-symmetry paths of magnon degeneracy in the altermagnet, the spin current behaviour retains its antiferromagnetic character, but for general modulation vectors $\bo{Q}$ away from these paths, the spin current response of the altermagnet is different from that of an antiferromagnet. In particular, the resonance frequencies becomes non-symmetric in frequency due to the broken magnon degeneracy. These results demonstrate that while the spin order in an altermagnet can give rise to THz spin pumping like for a regular antiferromagnet, the pumping behaviour can be qualitatively different depending on the relation between the spatial modulation of the magnetic field and the crystal orientation.

\begin{acknowledgments}
We thank B. Brekke and C. Sun for helpful dicussions. This work was supported by the Research
Council of Norway through Grant No. 323766 and its Centres
of Excellence funding scheme Grant No. 262633 “QuSpin.” Support from
Sigma2 - the National Infrastructure for High Performance
Computing and Data Storage in Norway, project NN9577K, is acknowledged. 
 \end{acknowledgments}

\bibliography{refs.bib}

\appendix

\begin{widetext}

\section{Interfacial exchange interaction}\label{sec:appA}
We consider the interaction term in more detail, considering the geometry depicted in Fig. \ref{fig: grid}.
\begin{equation}
\begin{split}
H_\text{int} &=-2\sum_{\langle i, j\rangle|i\in \text{AF}, j\in \text{NM} }J_i \begin{pmatrix}
\cc{j,\uparrow} & \cc[]{j, \downarrow}\end{pmatrix} \boldsymbol{\sigma}\begin{pmatrix}
\ca{j,\uparrow} \\ \ca{j, \downarrow} \end{pmatrix}\cdot \hat{\boldsymbol{S}}_i \\&= -2\sum_{\langle i, j \rangle \atop i\in A, j\in \text{NM}} J_A \big [S_{iA}^- \cc[]{j, \uparrow}\ca[]{j, \downarrow} + S_{iA}^+ \cc[]{j, \downarrow}\ca[]{j, \uparrow} \big]+ S_{iA}^z(\cc[]{j,\uparrow}\ca[]{j, \uparrow}- \cc[]{j, \downarrow}\ca[]{j, \downarrow})\\&\qquad\qquad-2\sum_{\langle i, j \rangle \atop i\in B, j\in \text{NM}} J_B \big [S_{iB}^- \cc[]{j, \uparrow}\ca[]{j, \downarrow} + S_{iB}^+ \cc[]{j, \downarrow}\ca[]{j, \uparrow} \big]+ S_{iB}^z(\cc[]{j,\uparrow}\ca[]{j, \uparrow}- \cc[]{j, \downarrow}\ca[]{j, \downarrow}) \\
&=-2\bigg(\sum_{\langle i, j \rangle \atop i\in A}J_A S_{iA}^+\cc[]{j,\downarrow}\ca[]{j, \uparrow} + \text{h.c.} + \sum_{\langle i, j\rangle \atop i \in B}J_B S_{iB}^+ \cc[]{j,\downarrow}\ca[]{j, \uparrow} + \text{h.c.} \bigg) + H_{int}^z
\end{split} \label{eqn: appA}
\end{equation}

where 
\begin{equation}
H_{int}^z=-2\sum_{\langle i,j\rangle \atop i\in A}S_{iA}^z(\cc[]{j,\uparrow}\ca[]{j, {\uparrow}}-\cc[]{j, \downarrow}\ca[]{j, \downarrow}) 
-2\sum_{\langle i,j \rangle \atop i\in B}S_{iB}^z(\cc[]{j,\uparrow}\ca[]{j, {\uparrow}}-\cc[]{j, \downarrow}\ca[]{j, \downarrow})
\end{equation}
Leaving $H_\text{int}^z$ aside for now, we shall denote denote the rest of the terms as $H_{int}^\parallel$. The term in $H_\text{int}^\parallel$ running over the \textit{A} sublattice becomes 

\begin{equation}
        \begin{split}
            -2\sum_{\langle i,j\rangle \atop i\in A} J_A S_{iA}^+ \cc[]{j, \downarrow}\ca[]{j, \uparrow}  & = -2\frac{\sqrt{2S}J_A}{N_N \sqrt{N_A}}\sum_{i\in A}\sum_{\boldsymbol{q}\in \lozenge}\sum_{\boldsymbol{k_1},\boldsymbol{k_2}\in\square} \aaa[]{\boldsymbol{q}}\cc[]{\boldsymbol{k_1}, \downarrow}\ca[]{\boldsymbol{k_2},\uparrow}e^{i\boldsymbol{q}\cdot\boldsymbol{r}_i}e^{i(\boldsymbol{k}_2 - \boldsymbol{k}_1)\cdot(\boldsymbol{r}_i + a\hat{\boldsymbol{z}})} \\
            &=-\frac{\overline{J}_A}{N_A}\sum_{i\in A}\sum_{\boldsymbol{q}\in \lozenge}\sum_{\boldsymbol{k_1},\boldsymbol{k_2}\in\square} \aaa[]{\boldsymbol{q}}\cc[]{\boldsymbol{k_1}, \downarrow}\ca[]{\boldsymbol{k_2},\uparrow}e^{i(\boldsymbol{k}_2 - \boldsymbol{k}_1 + \boldsymbol{q})\cdot \boldsymbol{r}_i} e^{ia(\boldsymbol{k}_2 - \boldsymbol{k}_1)\cdot \hat{\boldsymbol{z}}}
        \end{split}
\end{equation}
where $\overline{J}_A = 2\sqrt{2S}J_A N_A/(N_N\sqrt{N_A}$) and where we have assumed the spacing between the altermagnetic and normal-metal layer to be $a\hat{\boldsymbol{z}}$ where \textit{a} is the lattice parameter. The last exponential factor becomes unity as $\boldsymbol{k}_1, \boldsymbol{k}_2$ run in the \textit{xy} plane. In order to connect the Brillouin zone of the normal metal with the reduced Brillouin zone of the altermagnetic lattice, we rewrite the sum over the regular Brillouin zone as $\sum_{\boldsymbol{k}\in\square}f(\boldsymbol{k})=\sum_{\boldsymbol{k}\in\lozenge}[f(\boldsymbol{k})+f(\boldsymbol{k}+\boldsymbol{Q})]$ where $Q=\pi/a\hat{\boldsymbol{x}}+\pi/a\hat{\boldsymbol{y}}$ is the vector connecting the two. It then follows that
    \begin{equation}
        \begin{split}
        -\frac{\overline{J}_A}{N_A}\sum_{i\in A}\sum_{\boldsymbol{q}\in \lozenge}\sum_{\boldsymbol{k_1},\boldsymbol{k_2}\in\square} &\aaa[]{\boldsymbol{q}}\cc[]{\boldsymbol{k_1}, \downarrow}\ca[]{\boldsymbol{k_2},\uparrow}e^{i(\boldsymbol{k}_2 - \boldsymbol{k}_1 + \boldsymbol{q})\cdot \boldsymbol{r}_i} \\
        &=-\frac{\overline{J}_A}{N_A}\sum_{i\in A}\sum_{\boldsymbol{q}\in \lozenge}\sum_{\boldsymbol{k_1},\boldsymbol{k_2}\in\lozenge} \aaa[]{\boldsymbol{q}}e^{i\boldsymbol{q}\cdot \boldsymbol{r}_i}\big[(\cc[]{\boldsymbol{k}_1, \downarrow}e^{-i\boldsymbol{k}_1 \cdot \boldsymbol{r}_i}+\cc[]{\boldsymbol{k}_1^U, \downarrow}e^{-i\boldsymbol{k}_1^U \cdot \boldsymbol{r}_i})(\ca[]{\boldsymbol{k}_2, \uparrow}e^{i\boldsymbol{k}_2 \cdot \boldsymbol{r}_i}+\ca[]{\boldsymbol{k}^U, \downarrow}e^{i\boldsymbol{k}_2^U \cdot \boldsymbol{r}_i}) \big] \\
            &=-{\overline{J}_A}\sum_{\boldsymbol{q}\in \lozenge}\sum_{\boldsymbol{k}\in\square} \aaa[]{\boldsymbol{q}}\cc[]{\boldsymbol{k}, \downarrow}\ca[]{\boldsymbol{k}-\boldsymbol{q},\uparrow} + \kappa\aaa[]{\boldsymbol{q}}\cc[]{\boldsymbol{k}^U,\downarrow}\ca[]{\boldsymbol{k}-\boldsymbol{q}, \uparrow}
        \end{split}
    \end{equation}
where we have defined the Umklapp momentum $\boldsymbol{k}^U = \boldsymbol{k}+\boldsymbol{Q}$ and where $\kappa = 1$ for $\boldsymbol{r}_i \in A$ and $\kappa = -1$ for $\boldsymbol{r}_i \in B$, arising from a factor $e^{i\boldsymbol{Q}\cdot \boldsymbol{r}_i}$ in the cross terms. Performing the analogous calulation for the \textit{B} sublattice gives the final interaction term, 
    \begin{equation}
        \begin{split}
            H_{int}^\parallel=-\sum_{\boldsymbol{q}\in \lozenge} \sum_{\boldsymbol{k}\in \square} (M_{\boldsymbol{q}}^{\alpha R}\aaaa[]{\boldsymbol{q}}+M_{\boldsymbol{q}}^{\beta^\dagger R}\bc[]{\boldsymbol{q}}) \cc[]{\boldsymbol{k},\downarrow}\ca[]{\boldsymbol{k}-\boldsymbol{q}, \uparrow} +(M_{\boldsymbol{q}}^{\alpha U}\aaaa[]{\boldsymbol{q}}+M_{\boldsymbol{q}}^{\beta^\dagger U}\bc[]{\boldsymbol{q}})\cc[]{\boldsymbol{k}^U,\downarrow}\ca[]{\boldsymbol{k}-\boldsymbol{q}, \uparrow} +\text{h.c.}
        \end{split}
    \end{equation}

where we have defined 
\begin{equation}
\begin{split}
        M_{\boldsymbol{q}}^{\alpha\kappa} &= (\overline{J}_A u_{\boldsymbol{q}}+\kappa\overline{J}_B v_{\boldsymbol{q}})  \\
    M_{\boldsymbol{q}}^{\beta^\dagger \kappa}&= (\overline{J}_A v_{\boldsymbol{q}}+\kappa\overline{J}_B u_{\boldsymbol{q}})
\end{split}
\end{equation}
where $\kappa= 1$ for the regular scattering process (\textit{R)} and $\kappa=-1$ for the Umklapp process (\textit{U}). Rewriting this finally as a sum over the operators $\{\alpha_{\boldsymbol{q}}, \beta_{\boldsymbol{q}}^\dagger\}$, we obtain
\begin{equation}
    H_\text{int}^\parallel = -\sum_{\boldsymbol{q}\in\lozenge}\sum_{\boldsymbol{k}\in \square}\sum_{\kappa \in \{R,U\}}\sum_{\nu \in \{\alpha, \beta^\dagger \}}M_{\boldsymbol{q}}^{\nu \kappa}\nu_{\boldsymbol{q}}\cc[]{\boldsymbol{k}^\kappa, \downarrow}\ca[]{\boldsymbol{k}-\boldsymbol{q},\uparrow} + \text{h.c.}
\end{equation}

We now consider $H_{int}^z$. To first order in magnon operators, the term becomes  

    \begin{equation}
        \begin{split}
        H_{int}^z &= -2\sum_{\langle i,j \rangle \atop i\in A} J_A S(\cc[]{j, \uparrow}\ca[]{j,\uparrow}-\cc[]{j,\downarrow}\ca[]{j, \downarrow}) -2\sum_{\langle i,j \rangle \atop i\in B}J_B (-S)(\cc[]{j, \uparrow}\ca[]{j,\uparrow}-\cc[]{j,\downarrow}\ca[]{j, \downarrow}) \\
        &=-\frac{2J_A S}{N_N \sqrt{N_A}}\sum_{i\in A}\sum_{\boldsymbol{k}_1, \boldsymbol{k}_2 \in \square}(\cc[]{\boldsymbol{k}_1, \uparrow}\ca[]{\boldsymbol{k}_2, \uparrow}-\cc[]{\boldsymbol{k}_1, \downarrow}\ca[]{\boldsymbol{k}_2, \downarrow})e^{-i(\boldsymbol{k}_1 - \boldsymbol{k}_2)\cdot (\boldsymbol{r}_i + a\hat{\boldsymbol{z}})}-\frac{2J_B S}{N_N \sqrt{N_B}}\sum_{i\in B}\sum_{\boldsymbol{k}_1, \boldsymbol{k}_2 \in \square}(\cc[]{\boldsymbol{k}_1, \uparrow}\ca[]{\boldsymbol{k}_2, \uparrow}-\cc[]{\boldsymbol{k}_1, \downarrow}\ca[]{\boldsymbol{k}_2, \downarrow})e^{-i(\boldsymbol{k}_1 - \boldsymbol{k}_2)\cdot (\boldsymbol{r}_i + a\hat{\boldsymbol{z}})} \\
        &=-\sqrt{2S}\sum_{\boldsymbol{k}\in\square}(\overline{J}_A - \overline{J}_B)(\cc[]{\boldsymbol{k},\uparrow}\ca[]{\boldsymbol{k},\uparrow}-\cc[]{\boldsymbol{k},\downarrow}\ca[]{\boldsymbol{k},\downarrow}) + (\overline{J}_A + \overline{J}_B)(\cc[]{\boldsymbol{k}+\boldsymbol{Q},\uparrow}\ca[]{\boldsymbol{k},\uparrow}-\cc[]{\boldsymbol{k}+\boldsymbol{Q},\downarrow}\ca[]{\boldsymbol{k},\downarrow}) \\
        &=-\sqrt{2S}\sum_{\boldsymbol{k}\in \square}\sum_{\kappa\in\{R,U\}}(\overline{J}_A - \kappa\overline{J}_B)\big(\cc[]{\boldsymbol{k}^\kappa, \uparrow}\ca[]{\boldsymbol{k},\uparrow}-\cc[]{\boldsymbol{k}^\kappa, \downarrow}\ca[]{\boldsymbol{k}, \downarrow}\big)
        \end{split}
    \end{equation}

\section{Spin current derivation}\label{sec:appB}
We treat the spin current lesser Green's function in the interaction picture with the exchange interaction at the interface as a perturbation in the Keldysh formalism. The starting point is the lesser component of the contour-ordered Green's function for a given set of parameters $\nu$, $\boldsymbol{q}$, $\kappa$ (these indices are omitted in the following for brevity of notation), 
\begin{equation}
    G_\mathcal{C}(t_1, t_2)=G_\mathcal{C}^<(t_1, t_2)=-i\langle  M_{\boldsymbol{q}}^{\nu\kappa}\nu_{\boldsymbol{q}}^{H}(t_1)s_{\boldsymbol{q}}^{\kappa,-,{H}}(t_2)\rangle
\end{equation}
where $\mathcal{T}_\mathcal{C}$ is the contour-ordering operator placing operators with time arguments later on the contour to the left and earlier on the right. We now go to the interaction picture where we treat $H_\text{int}^\parallel$ as a perturbation. It then follows that the Green's function may be written as 
\begin{equation}
    G_\mathcal{C}^<(t_1, t_2)=-iM_{\boldsymbol{q}}^{\nu\kappa}\big\langle T_\mathcal{C}  \nu_{\boldsymbol{q}}(t_1)s_{\boldsymbol{q}}^{\kappa,-}(t_2) e^{-i\int_\mathcal{C}dt H_\text{int}^\parallel(t)} \big\rangle _0 
\end{equation}
Expanding the exponential to first order in $H_\text{int}^\parallel$, we obtain
\begin{align}
        G_\mathcal{C}^<(t_1, t_2) - G_\mathcal{C}^{<,0}(t_1, t_2) &\simeq -\big\langle \mathcal{T}_\mathcal{C}\int_\mathcal{C}dt \sum_{\boldsymbol{q}'\in\lozenge}\sum_{\kappa'\in\{R,U\}}\sum_{\nu'\in\{\alpha,\beta^\dagger\}} M_{\boldsymbol{q}}^{\nu\kappa}(M_{\boldsymbol{q}'}^{\nu'\kappa'})^* \nu_{\boldsymbol{q}}(t_1)s_{\boldsymbol{q}}^{\kappa,-}(t_2){\nu'}_{\boldsymbol{q'}}^\dagger (t)s_{\boldsymbol{q}'}^{\kappa',+}(t)\big\rangle_0  + \mathcal{O}(H_\text{int}^2) \\
        &=-\sum_{\boldsymbol{q}'\in\lozenge}\sum_{\kappa'\in\{R,U\}}\sum_{\nu'\in\{\alpha,\beta^\dagger\}}M_{\boldsymbol{q}}^{\nu\kappa}(M_{\boldsymbol{q}'}^{\nu'\kappa'})^*\int_\mathcal{C}dt \langle T_\mathcal{C}\nu_{\boldsymbol{q}}(t_1){\nu'}_{\boldsymbol{q'}}^\dagger (t)\rangle_0 \langle T_\mathcal{C}s_{\boldsymbol{q}}^{\kappa,-}(t_2)s_{\boldsymbol{q}'}^{\kappa',+}(t) \rangle_0   \\
        &= \sum_{\kappa'\in\{R,U\}}M_{\boldsymbol{q}}^{\nu\kappa}(M_{\boldsymbol{q}}^{\nu\kappa'})^* \big[G_{\nu}(\boldsymbol{q}) \bullet G_{s^+}^{\kappa\kappa'}(\boldsymbol{q}) \big](t_1, t_2) \label{eqn: figs}
\end{align}
where $s_{\boldsymbol{q}}^{\kappa,+}=(s_{\boldsymbol{q}}^{\kappa,-})^\dagger$ and we defined the Green's functions 
\begin{align}
    G_{\nu}(\boldsymbol{q}; t, t') &= -i\langle T_\mathcal{C} \nu_{\boldsymbol{q}}(t){\nu'}_{\boldsymbol{q}}^\dagger (t') \rangle_0 \\
    G_{s^+}^{\kappa\kappa'}(\boldsymbol{q}; t, t') &= -i\langle T_\mathcal{C}s_{\boldsymbol{q}}^{\kappa',+}(t) s_{\boldsymbol{q}}^{\kappa,-}(t') \rangle_0
\end{align}
and where the bullet product $\bullet$ denotes integration of the internal time parameter $t$ along the contour and is defined by 
 \begin{equation}
     (A\bullet B)(t_1, t_2) = \int_\mathcal{C}dt A(t_1, t)B(t, t_2)
 \end{equation}
Note that in Eq. (\ref{eqn: figs}), we have used that the Hamiltonian is diagonal in magnon operators $\alpha$, $\beta^\dagger$ as well as magnon momentum $\boldsymbol{q}$ to eliminate two of the summations. As indicated by the $<$ superscript, we consider the lesser component of the contour-ordered Green's function $G_\mathcal{C}(t_1, t_2)$. We proceed by utilizing the Langreth rules \cite{stefanucci2013nonequilibrium}. If 
\begin{subequations}
    \begin{align}
        C(t_1, t_2) &= (A\bullet B)(t_1, t_2) \\
        D(t_1, t_2) &= (A\bullet B \bullet C)(t_1, t_2)
    \end{align}
\end{subequations}
the advanced, retarded and lesser components of $C$ and $D$ satisfy 
\begin{subequations}
    \begin{align}
        C^< &= A^R\circ B^< + A^< \circ B^A \\
C^{R/A} &= A^{R/A} \circ B^{R/A} \\ 
D^< &= A^R \circ B^R \circ C^< + A^R \circ B^< \circ C^A + A^< \circ B^A \circ C^A \\
D^{R/A} &= A^{R/A}\circ B^{R/A} \circ C^{R/A}
    \end{align}
\end{subequations}
where we have defined the circle product 
\begin{equation}
    (A\circ B)(t_1, t_2) = \int_{-\infty}^\infty dt A(t_1, t)B(t, t_2)
\end{equation}
From this, it then follows that 
\begin{equation}
    G_\mathcal{C}^<(t_1, t_2)=\sum_{\kappa'\in\{R,U\}}M_{\boldsymbol{q}}^{\nu\kappa}(M_{\boldsymbol{q}}^{\nu\kappa'})^*\big[G_{\nu}^R(\boldsymbol{q})\circ G_{s^+}^{\kappa\kappa', <}(\boldsymbol{q}) + G_{\nu}^<(\boldsymbol{q})\circ G_{s^+}^{\kappa\kappa', A} (\boldsymbol{q})\big](t_1,t_2)
\end{equation}
The time integration reduces to a regular convolution integral as $G_\psi^R(t_1, t_2)$ and $G_\psi^<(t_1, t_2)$ depend only on the relative time $t_1-t_2$. The lesser component then becomes, upon a Fourier transformation of the Green's functions, a simple product.  Since we need $G^<_\mathcal{C}$ at equal times to compute the spin current, we now set $t_1=t_2=t$.
%Anticipating the final, equal-time result, we shall now set $t_2=t_1=t$ where $t_2$ is infinitesimally later than $t_1$. 
Considering then  for example the first term of $G_\mathcal{C}^<$, it follows, omitting momentum indices for brevity, that 

\begin{equation}
    \begin{split}
        &G_\nu^R\circ G_{s^+}^{\kappa\kappa',<}(t,t)\\&=\int_{-\infty}^\infty dt' G_\nu^R(t, t')G_{s^+}^{\kappa\kappa',<} (t', t) \\ 
        &=\int_{-\infty}^\infty dt'\int \frac{d\omega}{2\pi}\frac{d\omega'}{2\pi}e^{i\omega(t-t')}e^{i\omega'(t'-t)}G_\nu^R(\omega)G_{s^+}^{\kappa\kappa',<}(\omega') \\
        &=\int \frac{d\omega}{2\pi}G_\nu^R(\omega)G_{s^+}^{\kappa\kappa',<}(\omega)
    \end{split}
\end{equation}

The final expression for the correction to the lesser component of $G_{\boldsymbol{q},\kappa,\nu}^<(t,t)$ from Eq. (\ref{eqn: spin current commutator}), found by using the contour-ordered Green's function $G_\mathcal{C}^<$, is then given by 
\begin{equation}
    G_{\boldsymbol{q},\kappa,\nu}^<(t,t)-G_{\boldsymbol{q},\kappa,\nu}^{<,0}(t,t)=\int\frac{d\omega}{2\pi}\sum_{\kappa'\in\{R,U\}}M_{\boldsymbol{q}}^{\nu\kappa}(M_{\boldsymbol{q}}^{\nu\kappa'})^*\big[G_{\nu}^R(\boldsymbol{q},\omega)G_{s^+}^{\kappa\kappa', <}(\boldsymbol{q},\omega) + G_{\nu}^<(\boldsymbol{q}, \omega)G_{s^+}^{\kappa\kappa', A} (\boldsymbol{q},\omega) \big]
\end{equation}

\section{Spin pumping in the altermagnet}\label{sec:appC}
We now introduce an explicit time-dependence, $h^\pm (t)=h_0 e^{\mp i\Omega t}$ where $\Omega$ is the field frequency. We now want to treat Eq. (\ref{eqn: interaction}) as a perturbation,
\begin{equation}
    V=-\sum_{\pm}\lambda_{\pm\bo{Q}}\bigg[\big(\alpha_{\pm\bo{Q}}^{\vphantom{\dagger}} + \beta_{\pm\bo{Q}}^\dagger \big)h^-(t) + \big(\alpha_{\pm\bo{Q}}^{{\dagger}} + \beta_{\pm\bo{Q}}^{\vphantom{\dagger}} \big)h^+(t) \bigg]
\end{equation}
with $\lambda_{\pm\bo{Q}}=\zeta \frac{\sqrt{NS}}{4}(u_{\pm\bo{Q}} + v_{\pm\bo{Q}})$. As previously, we are interested in the magnon Green's functions $G_\nu^R$ and $G_\nu^<$ for $\nu\in\{\alpha, \beta^\dagger\}$ and their corrections due to the external time-dependent pumping field. In order to do this, we consider the contour-ordered Green's function for the magnon $\nu$ which to second order in $V$ becomes
\begin{align}
    G_{\nu,\bo{q}}(t_1,t_2)&=-i\big \langle \mathcal{T}_\mathcal{C } \nu_{\bo{q}}(t_1) \nu_{\bo{q}}^\dagger (t_2) e^{-i\int_\mathcal{C} dt V (t)}\big\rangle_0 \\ 
    &=-i\big \langle \mathcal{T}_c \nu_{\bo{q}}(t_1) \nu_{\bo{q}}^\dagger (t_2)\big\rangle_0 - \bigg\langle\mathcal{T}_\mathcal{C}\int_\mathcal{C}dt \nu_{\bo{q}}(t_1)\nu_{\bo{q}}^\dagger (t_2) V (t)\bigg\rangle_0+i\bigg\langle\mathcal{T}_\mathcal{C}\int_\mathcal{C} dtdt'\nu_{\bo{q}}(t_1)\nu_{\bo{q}}^\dagger (t_2) V (t)V (t')\bigg\rangle_0 \label{eqn: C2}
\end{align}
Here the $\langle ...\rangle_0$ signifies an average taken in the absence of $V$. The first-order term is clearly zero as it is odd in magnon operators. We consider the second-order term more closely. 
\begin{align}
    i\bigg\langle\mathcal{T}&_\mathcal{C}\int_\mathcal{C} dtdt'\nu_{\bo{q}}(t_1)\nu_{\bo{q}}^\dagger (t_2) V (t)V (t')\bigg\rangle_0\\&=2i\sum_{\pm }\sum_{\pm'}\lambda_{\pm\bo{Q}}\lambda_{\pm'\bo{Q}}\bigg\langle \mathcal{T}_\mathcal{C} \int_\mathcal{C}dtdt' h^+(t')h^-(t)\big[ \nu_{\bo{q}}(t_1)\nu_{\bo{q}}^\dagger (t_2) \nu_{\pm{\bo{Q}}}(t)\nu_{\pm'{\bo{Q}}}^\dagger (t')\big]\bigg\rangle \\ 
    &=2i\sum_{\pm}\sum_{\pm' }\lambda_{\pm\bo{Q}}\lambda_{\pm'\bo{Q}}\int_\mathcal{C}dt dt' h^+(t')h^-(t)\bigg[\big\langle \mathcal{T}_\mathcal{C} \nu_{\bo{q}}(t_1)\nu_{\pm'\bo{Q}}^\dagger (t') \big\rangle_0\big\langle \mathcal{T}_\mathcal{C} \nu_{\pm\bo{Q}}(t)\nu_{\bo{q}}^\dagger (t_2) \big\rangle_0 \label{eqn: 20} \\
    &\qquad\qquad\qquad\qquad + \big\langle \mathcal{T}_\mathcal{C} \nu_{\bo{q}}(t_1)\nu_{\bo{q}}^\dagger (t_2) \big\rangle_0\big\langle \mathcal{T}_\mathcal{C} \nu_{\pm\bo{Q}}(t)\nu_{\pm'\bo{Q}}^\dagger (t') \big\rangle_0 \bigg] \notag
\end{align}
where we have done a Wick expansion and where the second term in the expansion is zero. We have also used extensively that the Hamiltonian in absence of $V$ is diagonal in $\alpha,\beta^\dagger$ which allows us to discard several terms arising from the product $V^2$. The fact that the second term is zero can be seen as follows: Let us first define the quantity
\begin{equation}
    \Sigma(t_1, t_2)=\big\langle \mathcal{T}_\mathcal{C} h^+(t_1) h^-(t_2) \big \rangle
\end{equation}
We may then rewrite the second term in Eq. (\ref{eqn: 20}) as 

\begin{align}
    2i\sum_{\pm }\sum_{\pm' }\lambda_{\pm\bo{Q}}\lambda_{\pm'\bo{Q}}&G_{\nu,\bo{q}}^0(t_1, t_2)\int_\mathcal{C}dt dt' \Sigma(t', t) G_{\nu,\pm\bo{Q}}^0(t, t')\delta_{\pm, \pm'} \\
    &=-2i \sum_{\pm }\lambda_{\pm\bo{Q}}^2 G_{\nu,\bo{q}}^0(t_1, t_2)\int_\mathcal{C}dt' \big(\Sigma \bullet G_{\nu, \pm\bo{Q}} ^0 \big)(t',t') \\
    &=-2i \sum_{\pm}\lambda_{\pm\bo{Q}}^2 G_{\nu,\bo{q}}^0(t_1, t_2)\bigg(\int_{-\infty}^\infty dt' + \int_{\infty}^{-\infty}dt' \bigg) \big(\Sigma \bullet G_{\nu, \pm\bo{Q}} ^0 \big)(t',t') = 0 \label{eqn: C10}
\end{align}

Note that when multiplying out the term quadratic in the interaction $V$ in Eq. (\ref{eqn: C2}), we end up also with terms containing two $\alpha$ and two $\beta$ magnon operators such that the Green's function correction in principle could depend on both kinds of magnons. It however also follows that since such terms have the same time-structure as the second term in the Wick expansion, these terms are zero for the same reason as Eq. (\ref{eqn: C10}). 

We focus our attention now on the first term in the expansion and subtract the bare Green's function, 
\begin{align}
    \Delta G_{\nu,\bo{q}}^0 (t_1, t_2)&=-2i\sum_{\pm}\sum_{\pm'}\lambda_{\pm\bo{Q}}\lambda_{\pm'\bo{Q}} \int_\mathcal{C}dtdt' \Sigma(t',t)G_{\nu,\pm'\bo{Q}}^0(t_1, t')G_{\nu,\pm\bo{Q}}^0(t, t_2)\delta_{\bo{q}, \pm\bo{Q}}\delta_{\bo{q}, \pm'\bo{Q}}  \\
    &=-2i\sum_{\pm }\delta_{\bo{q}, \pm\bo{Q}}\lambda_{\pm\bo{Q}}^2 \big(G_{\nu, \pm\bo{Q}}^0\bullet \Sigma \bullet G_{\nu, \pm\bo{Q}}^0\big)(t_1, t_2) 
\end{align}
In order to regain the real-time Green's functions, we utilize once again the Langreth rules. The commutator of the self-energy $\Sigma(t_1, t_2)$ is zero as it contains no operators and thus, $\Sigma^\text{A/R}=0$ and $\Sigma^<=\Sigma$. It then follows that only the lesser component of the correction survives
\begin{equation}
    \Delta G_{\nu, \bo{q}}^<(t_1, t_2)=-2i\sum_{\pm} \delta_{\bo{q}, \pm\bo{Q}}\lambda_{\pm\bo{Q}}^2 \big(G_{\nu, \pm\bo{Q}}^R\circ \Sigma \circ G_{\nu, \pm\bo{Q}}^A\big)(t_1, t_2) 
\end{equation}
where we have defined the circle product 
\begin{equation}
    (A\circ B)(t_1, t_2) = \int_{-\infty}^\infty dt A(t_1, t)B(t, t_2)
\end{equation}

Using now that $h^\pm(t)=e^{\mp i\Omega t} \Rightarrow \Sigma(t_1,t_2)=h_0^2 e^{-i\Omega(t_1-t_2)}$. As both $\Sigma$ and $G_\nu^0$ depend only on relative time, the circle product reduces to a regular convolution. 

The Fourier transform of $\Sigma$ is 
\begin{equation}
    \Sigma(\omega)=\int d(t_1-t_2)e^{i\omega(t_1-t_2)}e^{-i\Omega(t_1-t_2)}h_0^2=2\pi h_0^2 \delta(\omega-\Omega)
\end{equation}
The second order correction then becomes 
\begin{equation}
    \Delta G_{\nu,\bo{q}}^< (\omega)=-4\pi i h_0^2 \delta(\omega-\Omega) \sum_{\pm}\lambda_{\pm\bo{Q}}^2 |G_\nu^R(\omega, \pm\bo{Q})|^2 \delta_{\bo{q}, \pm\bo{Q}}
\end{equation}

If we now consider the definition of the distribution function
\begin{equation}
    f^{\nu}(\omega, \bo{q}) = \frac{G_\nu ^<(\omega, \bo{k})}{2i \text{Im}\{G_\nu^R (\omega, \bo{k}) \}}
\end{equation}
and that to second order in the perturbation, only the lesser component of the correction survives, the distribution function is modified as follows, 
\begin{align}
    f^\nu(\omega, \bo{q}) &= n_B (\omega, T) -\frac{4\pi i h_0 ^2}{2i}\delta(\omega-\Omega)\bigg(\frac{-\eta^\nu}{(\omega - \omega_{\bo{q}}^\nu)^2 +(\eta^\nu)^2}\bigg)^{-1}\sum_{\pm}\lambda_{\pm\bo{Q}}^2 \frac{\delta_{\bo{q}, \pm\bo{Q}}}{(\omega-\omega_{\pm\bo{Q}}^\nu)^2 + (\eta^\nu)^2} \\ 
    &= n_B(\omega, T) + \frac{2\pi h_0^2}{\eta^\nu}\delta(\omega-\Omega)\sum_{\pm }\lambda_{\pm\bo{Q}}^2 \delta_{\bo{q}, \pm\bo{Q}}
\end{align}

\section{Spin susceptibility}\label{sec:appD}
In order to evaluate the spin current in Eq. (\ref{eqn: spin_current_green}), we need the retarded component of the spin susceptibility Green's function $G_{s^+}^R$ which follows from the imaginary time Matsubara Green's function 
\begin{equation}
    \overline{G}_{s^+}^{\kappa\kappa'}(\tau_1, \tau_2, \boldsymbol{q})=-\langle \mathcal{T}_\tau s_{\boldsymbol{q}}^{\boldsymbol{\kappa'+}}(\tau_1)s_{\boldsymbol{q}}^{\kappa -}(\tau_2) \rangle_{0 } \label{eqn: Matsubara Greens }
\end{equation}
through the analytical continuation 
\begin{equation}
    G_{s^+}^{\kappa\kappa', R} (\omega, \boldsymbol{q})=\overline{G}_{s^+}^{\kappa\kappa'}(i\omega_n \rightarrow \omega+i\eta, \boldsymbol{q})
\end{equation}
The Fourier transformed Matsubara Green's function is defined by 
\begin{equation}
    \overline{G}_{s^+}^{\kappa\kappa'}(i\omega_n, \boldsymbol{q})=\int_0^\beta d(\tau_1-\tau_2)\overline{G}_{s^+}^{\kappa\kappa'}({\tau_1, \tau_2, \boldsymbol{q}})e^{i\omega_n(\tau_1-\tau_2)}
\end{equation}
where $\omega_n=\frac{2\pi n}{\beta}$ are bosonic Matsubara frequencies.

We proceed by evaluating the imaginary time Green's function in Eq. (\ref{eqn: Matsubara Greens }) by inserting the explicit expression for the $s^{\pm}$ operators from Eq. (\ref{eqn: s_operator}). We also use that for a Hamiltonian diagonal in a quantum number $\nu$, e.g. the wavevector $\boldsymbol{k}$ as in the normal-metal Hamiltonian in Eq. (\ref{eqn: normal metal}) , the time-dependence of the operators $\cc[]{\boldsymbol{k},\sigma}$/$\ca[]{\boldsymbol{k},\sigma}$ follow from the Heisenberg equations and becomes 
\begin{subequations}
    \begin{align}
        \cc[]{\boldsymbol{k}, \sigma}(\tau) &= e^{\xi_{\boldsymbol{k}}\tau}\cc[]{\boldsymbol{k}, \sigma} \\
        \ca[]{\boldsymbol{k}, \sigma}(\tau) &= e^{-\xi_{\boldsymbol{k}}\tau}\cc[]{\boldsymbol{k}, \sigma}
    \end{align}
\end{subequations}
Using this, the imaginary time Green's function becomes 
    \begin{equation}
        \begin{split}
            \overline{G}_{s^+}(\tau_1, \tau_2, \boldsymbol{q})&=-\sum_{\boldsymbol{k}_1, \boldsymbol{k}_2}\langle {T}_\tau \cc[]{\boldsymbol{k}_1 -\boldsymbol{q}, \uparrow}(\tau_1)\ca[]{\boldsymbol{k}_1^{\kappa'}, \downarrow}(\tau_1)\cc[]{\boldsymbol{k}_2^\kappa, \downarrow}(\tau_2) \ca[]{\boldsymbol{k}_2 - \boldsymbol{q}, \uparrow}(\tau_2) \rangle_{{0}}  \\
            &=-\sum_{\boldsymbol{k}_1,\boldsymbol{k}_2} \big[ \langle{T}_\tau \cc[]{\boldsymbol{k}_1 -\boldsymbol{q}, \uparrow}(\tau_1)\ca[]{\boldsymbol{k}_1^{\kappa'}, \downarrow}(\tau_1) \rangle_{{0} }  \langle{T}_\tau \cc[]{\boldsymbol{k}_2^\kappa, \downarrow}(\tau_2)\ca[]{\boldsymbol{k}_2 - \boldsymbol{q}, \uparrow} (\tau_2)\rangle_{{0} }  - \langle {T}_\tau \mathcal{T}_\tau \cc[]{\boldsymbol{k}_1 -\boldsymbol{q}, \uparrow}(\tau_1) \ca[]{\boldsymbol{k}_2 - \boldsymbol{q}, \uparrow}(\tau_2)\rangle_{{0} }  \langle\mathcal{T}_\tau \cc[]{\boldsymbol{k}_2^\kappa , \downarrow}(\tau_2)\ca[]{\boldsymbol{k}_1^{\kappa'}, \downarrow}(\tau_1) \rangle_{{0} }  \big] \\
            &= \sum_{\boldsymbol{k}_1,\boldsymbol{k}_2}\big[\theta(\tau_1 - \tau_2)\langle  \cc[]{\boldsymbol{k}_1 -\boldsymbol{q}, \uparrow}\ca[]{\boldsymbol{k}_2 - \boldsymbol{q}, \uparrow}\rangle_{{0} }  \langle\ca[]{\boldsymbol{k}_1^{\kappa'}, \downarrow}\cc[]{\boldsymbol{k}_2^\kappa , \downarrow}\rangle_{{0} }  \\&\qquad\qquad\qquad+ \theta(\tau_2 - \tau_1)\langle \ca[]{\boldsymbol{k}_2 - \boldsymbol{q}, \uparrow}\cc[]{\boldsymbol{k}_1 -\boldsymbol{q}, \uparrow}\rangle_{{0} }  \langle\cc[]{\boldsymbol{k}_2^\kappa , \downarrow}\ca[]{\boldsymbol{k}_1^{\kappa'}, \downarrow}\rangle_{{0} } \big]
            e^{(\xi_{\boldsymbol{k}_1 - \boldsymbol{q}}-\xi_{\boldsymbol{k}_1^{\kappa'}})\tau_1}e^{-(\xi_{\boldsymbol{k}_2 - \boldsymbol{q}} - \xi_{\boldsymbol{k}_2^\kappa})\tau_2} \\
            &=\sum_{\boldsymbol{k}}\big[\theta(\tau_1-\tau_2)n_F(\xi_{\boldsymbol{k}-\boldsymbol{q}},T)(1-n_F(\xi_{\boldsymbol{k}^\kappa}, T))+\theta(\tau_2-\tau_1) n_F(\xi_{\boldsymbol{k}^\kappa}, T)(1-n_F(\xi_{\boldsymbol{k}-\boldsymbol{q}}, T))\big]e^{(\xi_{\boldsymbol{k} - \boldsymbol{q}}-\xi_{\boldsymbol{k}^\kappa})(\tau_1-\tau_2)}
        \end{split}
    \end{equation}
where $n_F$ is the Fermi-Dirac distribution for fermions ans \textit{T} is temperature. Note that in the second line above, the first term vanishes as the normal-metal Hamiltonian is spin-diagonal. We have also used that the normal-metal Hamiltonian is diagonal in momentum, giving rise to a $\delta_{\boldsymbol{k}_1, \boldsymbol{k}_2}$ and $\delta_{\kappa',\kappa}$ which simplifies the expression. We now continue by Fourier transforming the Green's function,
    \begin{equation}
        \begin{split}
            \overline{G}_{s^+}(i\omega_n, \boldsymbol{q})&=\int_0^\beta d(\tau_1-\tau_2)\overline{G}_{s^+}(\tau_1, \tau_2, \boldsymbol{q})e^{i\omega_n(\tau_1 - \tau_2)} \\
            &=-\sum_{\boldsymbol{k}}n_F(\boldsymbol{k}-\boldsymbol{q},T)(1-n_F(\boldsymbol{k^\kappa}, T))\int_0^\beta d(\tau_1 - \tau_2)e^{(i\omega_n + \xi_{\boldsymbol{k}-\boldsymbol{q}}-\xi_{\boldsymbol{k}^\kappa})(\tau_1- \tau_2) } \\
            &=-\sum_{\boldsymbol{k}}n_F(\xi_{\boldsymbol{k}-\boldsymbol{q}},T)(1-n_F(\xi_{\boldsymbol{k}^\kappa}, T)) \frac{e^{\beta(\xi_{\boldsymbol{k}-\boldsymbol{q}}-\xi_{\boldsymbol{k}^\kappa})} - 1}{i\omega_n + \xi_{\boldsymbol{k}-\boldsymbol{q}}-\xi_{\boldsymbol{k}^\kappa}} \\
            &=\sum_{\boldsymbol{k}}\frac{n_F(\xi_{\boldsymbol{k}-\boldsymbol{q}}, T)-n_F(\xi_{\boldsymbol{k}^\kappa}, T)}{i\omega_n +\xi_{\boldsymbol{k}-\boldsymbol{q}}-\xi_{\boldsymbol{k}^\kappa}}
        \end{split}
    \end{equation}

It then follows that the retarded spin susceptibility Green's function is given by \begin{equation}
    G_{s^+,\kappa\kappa'}^{R} (\omega, \boldsymbol{q})=\delta_{\kappa,\kappa'}G_{s^+, \kappa}^{R} (\omega, \boldsymbol{q})=\delta_{\kappa,\kappa'}\sum_{\boldsymbol{k}}\frac{n_F(\xi_{\boldsymbol{k}-\boldsymbol{q}}, T)-n_F(\xi_{\boldsymbol{k}^\kappa}, T)}{\omega +\xi_{\boldsymbol{k}-\boldsymbol{q}}-\xi_{\boldsymbol{k}}^\kappa+i\eta^N}
\end{equation}

\end{widetext}

\end{document}